\documentclass[aps,pra,preprint,superscriptaddress,floatfix]{revtex4-1}
\usepackage{amsmath,amsthm,amsfonts,amssymb,amscd}
\usepackage{fullpage}
\usepackage{lastpage}
\usepackage{graphicx}
\usepackage{wrapfig}
\usepackage{enumerate}
\usepackage{fancyhdr}
\usepackage{mathrsfs}
\usepackage{mathtools}
\usepackage{braket}
\usepackage{slashed}
\usepackage{bm}

\graphicspath{{./Figures/}{./nature_physFigures/}}

\begin{document}

\title{Active Tension Network model reveals an exotic mechanical state realized in epithelial tissues}
\author{Nicholas Noll}
\affiliation{Department of Physics, University of California Santa Barbara}
\author{Madhav Mani} 
\affiliation{Department of Applied Mathematics, Northwestern University }
\affiliation{Kavli Institute for Theoretical Physics}
\author{Idse Heemskerk} 
\affiliation{Department of Biosciences, Rice University}
\affiliation{Kavli Institute for Theoretical Physics}
\author{Sebastian Streichan} 
\affiliation{Kavli Institute for Theoretical Physics}
\author{Boris I. Shraiman}
\affiliation{Department of Physics, University of California Santa Barbara}
\affiliation{Kavli Institute for Theoretical Physics}

\begin{abstract}
Mechanical interactions play a crucial role in epithelial morphogenesis, yet understanding the complex mechanisms through which stress and deformation affect cell behavior remains an open problem. Here we formulate and analyze the Active Tension Network (ATN) model, which assumes that mechanical balance of cells is dominated by cortical tension and introduces tension dependent active remodeling of the cortex. We find that ATNs exhibit unusual mechanical properties: i) ATN behaves as a fluid at short times, but at long times it supports external tension, like a solid; ii) its mechanical equilibrium state has extensive degeneracy associated with a discrete conformal - ``isogonal" - deformation of cells. ATN model predicts a constraint on equilibrium cell geometry, which we demonstrate to hold in certain epithelial tissues. We further show that isogonal modes are observed in a fruit fly embryo, accounting for the striking variability of apical area of ventral cells and helping understand the early phase of gastrulation. Living matter realizes new and exotic mechanical states, understanding which helps understand biological phenomena.

\end{abstract}

\maketitle

Mechanics of growth and cellular rearrangements plays an important role in morphogenesis as both processes are central to defining the shape of developing tissues. As such, it has become a subject of intense study aiming to characterize specific mechanical processes involved in cell and tissue-wide dynamics\cite{Bell13,Lenne08,Julicher07,Wiescahus14}, uncover the regulatory mechanisms \cite{Lecuit07}, and identify if and how the mechanical state of the cell feeds back onto the larger developmental program \cite{Chen05,Shraiman05,Zallen09}. 


Epithelial tissue is a monolayer of apico-basally polarized cells tightly connected to their lateral neighbors \cite{Gilbert00}. Viewed from the apical side, cells form an approximately polygonal tiling of the plane. Mechanical integrity of the epithelial layer derives largely from the cortical actin-myosin network \cite{Levine08,Stamenovic02} localized as a planar ring just inside the cell's lateral surface \cite{Paluch12}. Each cell's cortical cytoskeleton is linked to those of the neighboring cells via cadherin-mediated adherens junctions \cite{Nelson08}. The equilibrium geometry of cells is determined by the balance of cytoskeletal and adhesive forces \cite{Lecuit07} within the tissue. Unlike passive materials, cells actively regulate such forces through mechano-transduction and internal remodeling, resulting in an intrinsically dynamic relation of stress and strain  and controllable plasticity \cite{Chen09, Weitz07}, which can drive rearrangement of cells. Elucidation of the manner in which cellular activity manifests in collective properties of the tissue is critical to understanding morphogenesis. 

Here we formulate a phenomenological model of an epithelial tissue as a two dimensional Active Tension Network (ATN), which in addition to cytoskeletal elasticity describes cytoskeletal re-arrangement through myosin activity and the recruitment of myosin into cytoskeletal fibers, thus capturing the plastic and adaptive response of cells to external stress. We shall explore static and dynamic properties of the ATN model, validate some of its predictions by comparing with experimental observations and identify new directions of further study.  

 
\subsection*{Formulation of the Active Tension Net Model}

Epithelial monolayers may be approximately represented by two-dimensional polygonal tilings, parameterized by the set of vertex coordinates $\{\bm{r}_i\}$ and are often described in terms by vertex models  \cite{Honda83,Julicher07} which assume that geometry of cells minimizes an elastic energy defined in terms of cell edge length ($r_{ij}=|\bm{r}_i-\bm{r}_j|$) and cell area ($A_{\alpha}$). We shall introduce a generalized class of vertex models by adding internal variables to capture active adaptation of the cytoskeleton. We begin by defining mechanical energy in its differential form \cite{Shraiman12}
\begin{equation}
dE[\{\bm{r}_i\}] = \displaystyle\sum\limits_{<i,j>} T_{ij} \, d r_{ij} + \displaystyle\sum\limits_{\alpha} p_\alpha \, dA_{\alpha}
\end{equation}
Here tension $T_{ij}$ defines the change in mechanical energy in response to the change of edge length by $d r_{ij}$ and
2D `apical pressure'  $p_\alpha$ defines the response to the change in cortical area by $dA_{\alpha}$. Tension Nets correspond to the situation when pressure differentials between cells are small so that mechanical balance is dominated by the tensions, in which case $p_{\alpha} \approx p_0$ with $p_0$ effectively controlling only the total area of cells and preventing the collapse of the array under the action of tension. 

Dynamics of vertex positions  is assumed to be relaxational and in the Tension Net approximation becomes
\begin{equation}
\nu \frac{d}{dt}\bm{r}_i = -\partial_{\bm{r}_i} E = \displaystyle\sum\limits_{\{j\}_i} T_{ij} \bm{\hat{r}}_{ij}
\end{equation}
where $\{j\}_i$ denotes the set of all vertices connected to vertex $i$, $\bm{\hat{r}}_{ij}$ is a unit vector in the direction from $\bm{{r}}_{i}$ to $\bm{{r}}_{j}$ and $\nu$ represents the effective friction between apical cytoskeleton and its substrate \cite{Simha13}, which sets the timescale of mechanical relaxation. 
Mechanical equilibrium of a Tension Net is reached when tensions balance at each vertex:  the right hand side of Eq. (2) is zero. Geometrically this corresponds to the three vectors ${\bm T}_{ij}=T_{ij} \bm{\hat{r}}_{ij}$ making up a triangle and since adjacent vertices share an edge, global tension balance means that  the set of $T_{ij}$'s defines a triangulation  \cite{Maxwell64, Chakraborty07} (see Fig. 1a,b).

Edge tension $T_{ij}$ depends on the  edge length $r_{ij}$ and intrinsic variables representing local state of the actomyosin bundle and cadherin mediated adhesion between cells. We shall  start with a particularly simple form, $T_{ij}=k(r_{ij}-\ell_{ij})$, parameterizing the internal state of the each interface by an intrinsic ``rest length" $\ell_{ij}$ of the underlying actomyosin bundle. The latter has dynamics of its own. Specifically, we shall assume that
\begin{equation}
\ell_{ij} ^{-1} \frac{d}{dt}\ell_{ij} = \tau_{\ell}^{-1} W\left( \frac{T_{ij}}{m_{ij} a T_s} \right)
\end{equation} 
The generic features of "walking kernel" $W(x)$, illustrated in Fig.1d, are known from single-molecule experiments \cite{Rief05,Rock10}: each myosin will walk contracting the actin bundle, unless the load per myosin, $T_{ij}/ a m_{ij}$, reaches the ``stall force" level $T_s$. (The characteristic length $a$  describes the extent to which myosin motors share mechanical load.)  Above this critical value, the filament simply elongates as each motor slips backwards \cite{Fisher07}.

Eqs. $(2,3)$ define the dynamics of a Tension Net with a specified myosin distribution on interfaces. The fixed point of these equations is then reached when i) tensions balance at all vertices and ii) all edges are at their stall tension, set by the local myosin level ($T_{ij}=aT_s m_{ij}$). Because tension balance requires the set of $T_{ij}$'s to form a triangulation, edge tensions, and hence myosin levels, cannot be prescribed independently. Luckily, in reality myosin levels are not fixed and are known to themselves respond to mechanical cues \cite{Zallen09,Poille09}, although the exact form of this mechanical feedback is not fully understood. Here we shall propose a particular form of mechanical feedback on myosin, that will ensure convergence to a balanced state. The latter is achieved if myosin recruitment depends on internal strain rate: 
\begin{equation}
m_{ij}^{-1} \frac{d}{dt}m_{ij} =  \alpha \ell_{ij}^{-1} \frac{d\ell_{ij}}{dt} = \alpha \tau_{\ell}^{-1} \, W\left( \frac{T_{ij}}{m_{ij} aT_s} \right)
\end{equation}
with $\alpha \ll 1$ parameterizing the rate of myosin recruitment which we assume to be slow relative to both mechanical relaxation and actomyosin contractility. This ``Dynamic Recruitment" form of mechanical feedback builds up myosin on overloaded and therefore ``slipping" bundles and reduces myosin on underloaded and contracting bundles until the stall condition is reached, bringing the system to equilibrium. This hypothesis is dictated by the requirement of ATN stability 
and 
should be regarded as a prediction of the model, to be tested by future experiments.

\subsection*{Equilibrium Manifold of a Tension Net}
\begin{figure}[h]
\centerline{
\includegraphics[width=.5\textwidth]{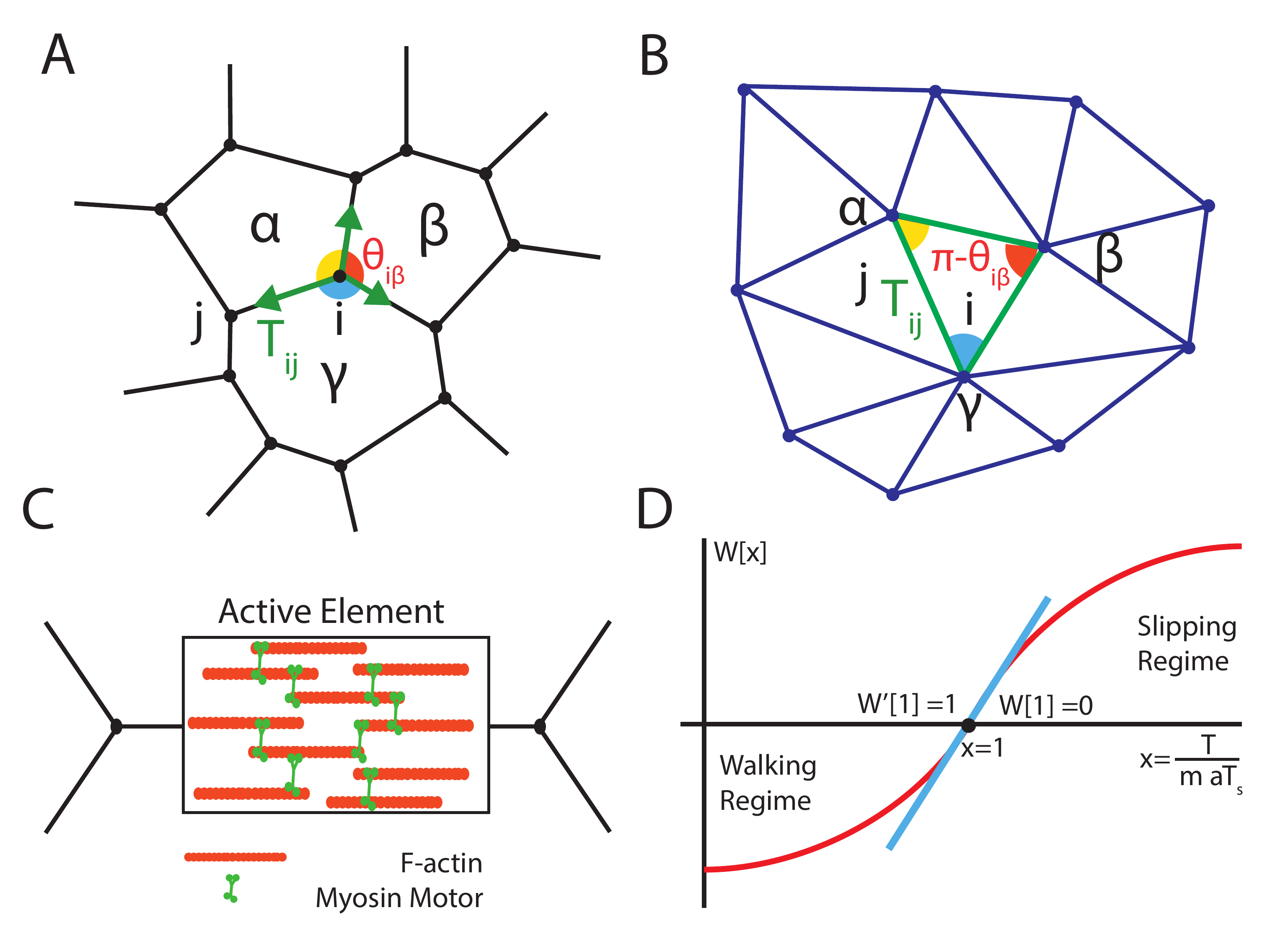}
}
\caption{(a) Tension Net representation of a 2D array of cells. In mechanical equilibrium tensions balance at each vertex; (b) Tensions corresponding to an equilibrium state form a triangulation, with triangle angles being supplementary to the angles at the corresponding vertex; (c) A cartoon representation of the actomyosin bundle as the active element of the cortical cytoskeleton (d) Dependence of the actomyosin bundle contraction rate to the mechanical load. The "walking kernel" $W(x)$, see Eq. (3), changes sign from contraction to elongation  when mechanical load per myosin $T/am$ exceed the stall load $T_s$. }
\end{figure}  
Mechanical equilibrium a Tension Net requires a local balance of tensions (see Eq. $(2)$) which relates the geometry of the cell array to a triangulation of the ``tension plane" (see Fig. 1). This relation involves the $\theta_{i\alpha}$ angles at vertex $i$ which define angles of the $i^{th}$ tension triangle with  $\pi-\theta_{i\alpha}$ being the angle opposite  of the $T_{ij}$ side (see Fig.1).  The fact that triangles corresponding to vertices that belong to the the same cell  pack together into a triangulation constrains the angles
\begin{equation}
\chi_{\alpha}=\displaystyle\prod\limits_{i\in\mathcal{V}(\alpha)} {\sin \theta_{i,\alpha} \over \sin \theta_{i-1,\alpha}}=1
\end{equation}
where $i$ labels the vertices of cell $\alpha$, denoted as $\mathcal{V}(\alpha)$, in a clockwise fashion. This constraint follows directly from the sine law applied to the triangles that share vertex $\alpha$ of the triangulation. A polygonal array with all $\chi_{\alpha}=1$ is geometrically $compatible$ with tension balance equilibrium. Since $\chi_{\alpha}$ can be readily computed for any polygonal array, the compatibility constraint allows one to asses whether a given cell array could be a balanced tension net.

To count the number of degrees of freedom that define balanced tension configurations
we note that a triangulation is completely specified by the positions of its vertices, 
the number of which equals $c$ - the number of polygonal cells - so that triangulation is specified by $2c$ independent degrees of freedom. 
Note that the number of edges $e = 3c$ is larger than $2c$ meaning that $T_{ij}$ cannot be chosen independently; the balanced set of tensions satisfies $c$ angular constraints imposed by the planarity of the triangulation. 


Next we observe that the number of degrees of freedom for equilibrium tension net geometries, given by $2v-c=3c$ ($v$ being the number of vertices of the cell array), is larger than the $2c$ degrees of freedom for the dual triangulations of the tension plane. Hence, a given set of tensions must correspond to multiple possible cell arrays: specifically, to a manifold of nets with one degree of freedom per cell.  
Given a triangulation we can construct a dual polygonal lattice by defining the circumcircle center for each triangle and drawing a Voronoi lattice based on these centroids. Yet, such a lattice is not the unique dual of the tension triangulation. As long as none of the vertex angles are perturbed, we can freely ``inflate" or ``deflate" lattice cells, as illustrated in Fig. 2(a),  with no cost of energy and thus without disturbing mechanical equilibrium and the underlying tension-triangulation. Quite generally such angle preserving - ``isogonal" -deformations have the form
\begin{equation}
\delta \bm {r}_i=S_{\alpha\beta\gamma}^{-1}[{\bm T}_{ij} \Theta_{\gamma}+{\bm T}_{ik} \Theta_{\alpha}+
{\bm T}_{il} \Theta_{\beta}]
\end{equation}
where $\delta \bm {r}_i$ denotes displacement of vertex at which adjacent cells $\alpha, \beta, \gamma$ meet and $S_{\alpha\beta\gamma}$  (Fig. 1ab) denotes the area of said vertex's dual triangular plaquette and $\Theta_{\alpha}$'s are independent parameters. Tensions $ T_{ij } $ appear as coefficients because their ratios capture the implicit geometric constraints within tension nets central to the structure of the isogonal modes. (Note for example that $\delta \bm{r}_{i}=0$ for $\Theta_{\alpha}=\Theta_{\beta}=\Theta_{\gamma}$.)
The compatibility condition (see Eq. 5) satisfied by equilibrium tension nets is essential for allowing such isogonal dilation modes to exist! Because they do not invoke a restoring force, isogonal deformations are the easily excitable "soft modes"
which are expected to dominate observed fluctuations of tension nets close to mechanical equilibrium. 

We note also that isogonal modes can be thought of as a discrete manifestation of the conformal symmetry that appears in 2D continuum elasticity in the case of a vanishing bulk modulus (see SI for details). Isogonal modes also generalize the isoperimetric ``breathing modes" of a hexagonal lattice \cite {Villain80}.

\begin{figure}[h]
\centerline{
\includegraphics[width=.5\textwidth]{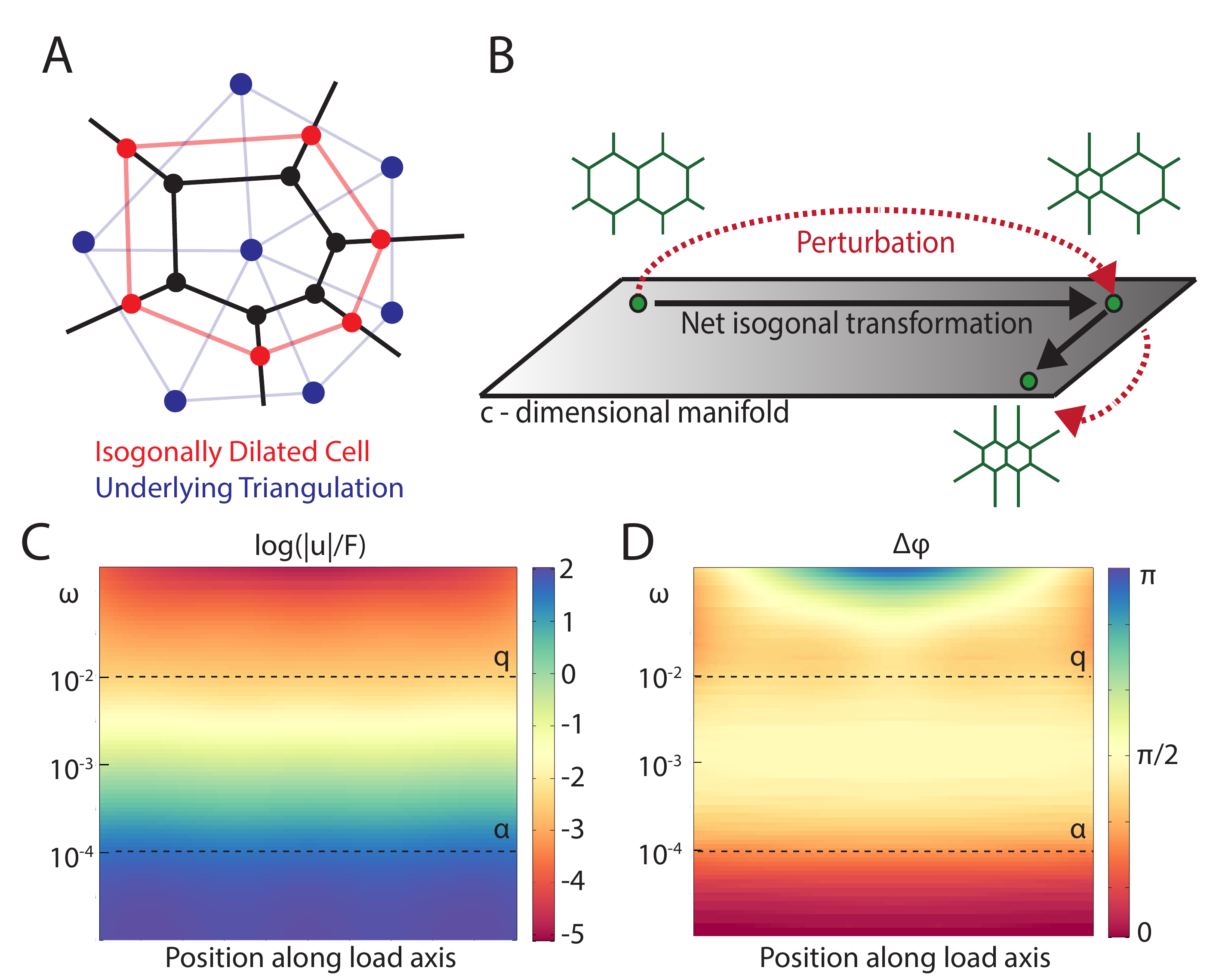}
}
\caption{(a) Cartoon illustrating the {\it isogonal}, i.e. angle preserving, `breathing mode'  of a cell in a tension net. (b)
Because ATN equilibrium is a manifold rather than a point, after a transient perturbation the system does not necessarily return to the same state, resulting in an `isogonal' transformation. (c) Amplitude and (d) phase of the strain (as a function of position in a 2D sheet) in response to periodic forcing $T_B \cos \omega t$ at the boundaries ( $\alpha = 10^{-2}$ and $\omega = 10^{-4}$). As the frequency decreases below  $\alpha$ the phase shifts from $0$ to $\pi/2$ indicating crossover from viscous fluid behavior to an elastic solid. Crossover at $\omega \sim q$ corresponds to viscoelasticity.}
\end{figure} 

\subsection*{Dynamical properties of Active Tension Nets.}
Let us  consider the dynamics of small perturbations around a mechanical equilibrium state, which can be described by linearizing Eqs. (2-4).
While detailed calculations are carried out in the SI, the key features can be understood from a vastly simpler analysis of a 1D ``active tension chain" model which has the form
\begin{eqnarray}
&{d \over dt} \delta  {T}_n=D \nabla^2  \delta T_{n}-q(\delta T_{n}-\delta m_n) \\
& {d \over dt} \delta  {m}_n=\alpha (\delta T_{n}-\delta m_n)
\end{eqnarray}
where $\delta T_n$ and $\delta m_n$ are deviations from the equilibrium state, $n$-is an integer indexing the edges along the chain, $ \nabla^2 $ is the discrete 1D Laplacian and $D,q,\alpha$ parameters are derived by linearization of Eqs. (2-4), see SI. The rate of strain in the chain is ${\dot u}_n ={d \over dt} ( {r}_{n+1}- r_n) =\nu^{-1} \nabla^2  \delta T_{n}$. With that, Eq. 7 is recognized as the Maxwell model of viscoelasticity forced by myosin. With a constant $\delta m$ forcing these would predict persistent flow (i.e. non-zero rate of strain ),  perturbations of tension being exponentially localized with characteristic ``screening length" $\lambda=\sqrt{D/q}$. At long times myosin recruitment (with $\alpha \ll q$) is important and the chain converges towards mechanical equilibrium: $\delta m_n \approx \delta T_{n}-q^{-1} D \nabla^2 T_n$ so that ${d \over dt} \delta  {m}_n \approx \alpha q^{-1} D \nabla^2 T_n$ and ${d \over dt} T_n \approx \alpha D \nu^{-1} q^{-1} {\dot u}_n$ (see SI) This means that unlike viscoelastic flow response at short time, the long time behavior is effectively elastic with $k_{eff} \sim \alpha \nu^{-1} q^{-1}D$.
A similar crossover from fluid-like response at intermediate time to solid-like behavior at long times occurs in the fully 2D ATN (see Fig. 2cd).
 
\subsection*{ATN predictions and the Ventral Furrow (VF) formation in $Drosophila$ embryo.}

One of the striking predictions of the ATN model is the existence of the isogonal soft modes that allow easy variability of cell area. Extreme variability of apical cell area has in fact been observed in the beginning of the gastrulation process in {\it Drosophila}, when cells along the ventral midline of the embryo constrict their apical surfaces, initiating the formation of a furrow that subsequently internalizes the future mesoderm \cite{Wieschaus91}, as shown in Fig. 3(ab). This apical constriction was shown to be driven by pulsed contractions of a {\it medial} actomyosin network (located on the apical cell surface) that connects to the adherens junction-anchored cortical cytoskeleton. The process has been described as a ``ratchet" \cite{Wieschaus09} where medial myosin pulses cause transient constrictions, that are subsequently stabilized by the retracted cytoskeletal cortex.  

This phenomenon is readily interpretable in terms of the ATN model. If we assume that the {\it cortical} myosin concentrations are relatively static over the timescale of medial myosin pulsing, the ATN model predicts that any transient perturbation of mechanical balance due to medial myosin contractions would leave behind an isogonal deformation of the cell array, as it returns to mechanical balance dominated by cortical tensions that remain unchanged. Hence we predict that cell deformation during the early stages of ventral furrow formation should be well described by motion along an isogonal manifold.

However, before testing this prediction, we can test the applicability of the tension balance hypothesis that underlies the ATN model. While it is not yet possible to measure all internal tensions in a live tissue, the compatibility constraint (3) provides us with an indirect way to evaluate if tension balance may be playing a role in defining geometry of cells. To that end we examine apical snapshots of  tissues (e.g. Fig 3a) and calculate $\chi_\alpha$  for each cell. We then compare the resultant probability distribution function (PDF) of $\log \chi$ to a  ``null"  distribution constructed  from a fictitious cell array with the angles reshuffled while preserving their empirical distribution (see SI for more details). A tension net close to mechanical equilibrium should generate a PDF of $\chi_\alpha$ clustering significantly closer to $\chi=1$ than the null distribution.  Fig. 3(c) presents the result of such an analysis for the embryonic mesoderm. Based on the analysis of $\sim$ 5000 cells, we find strong and highly statistically significant (Kolmogorov-Smirnov \cite{Massey51} $p<10^{-9}$) accumulation of $\log \chi_{\alpha}$ near zero, consistent with approximate tension balance. (Note that one cannot expect the compatibility constraint to be exact on account of cellular  fluctuations and of the noise introduced in image analysis.) This finding is non-trivial as similar analysis of cells in wing imaginal discs \cite{Irvine14} reveals no statistically significant tendency towards $\log \chi_{\alpha} \approx 0$, from which we conclude that tension balance is not a good description of imaginal disc epithelia. (Yet other tissues that we have analyzed (see SI) offer more examples of applicability of the ATN model.)

 
\begin{figure}[h]
\centerline{
\includegraphics[width=.5\textwidth]{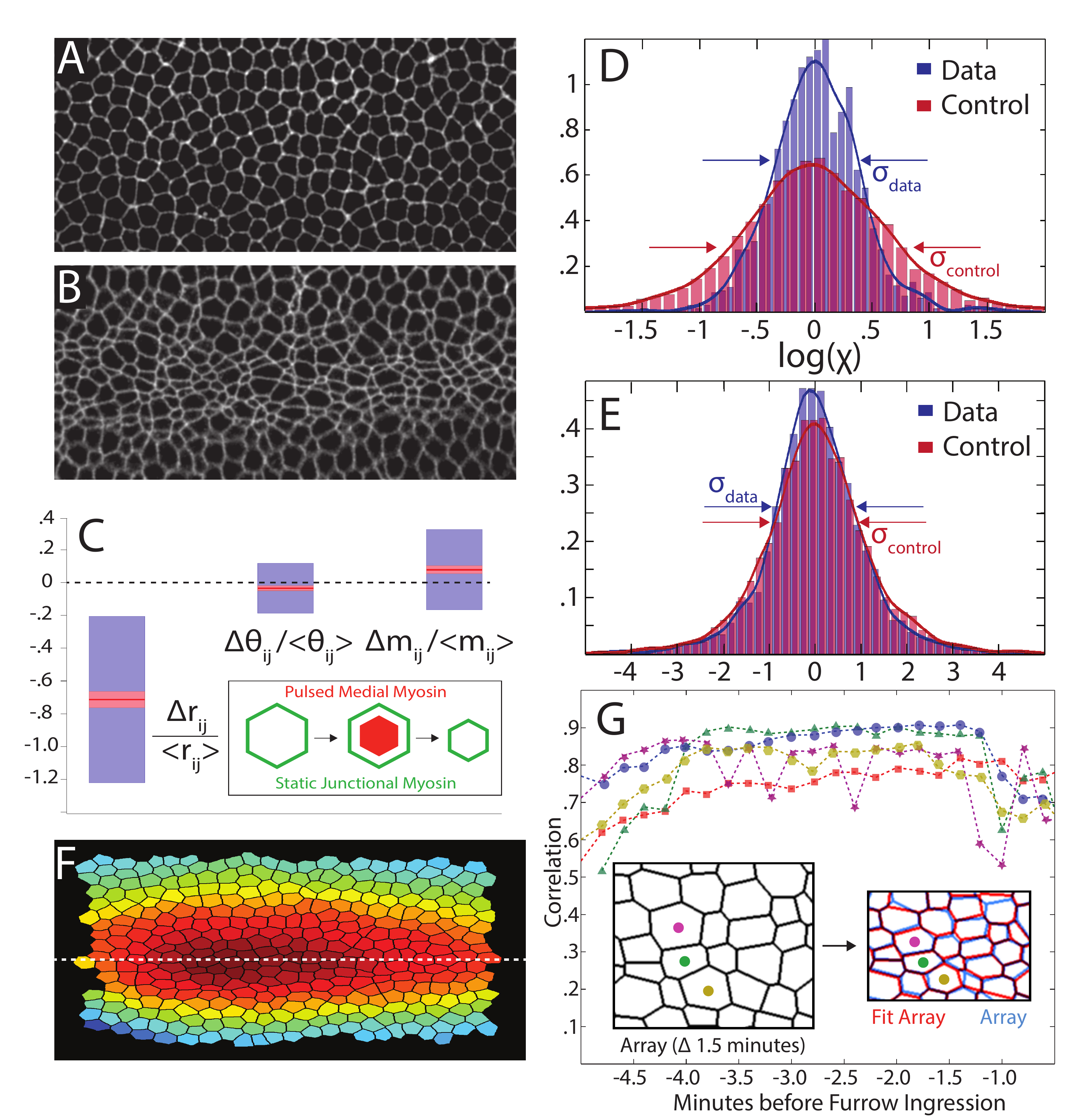}
}
\caption{(a) A ventral view of \emph{Drosophila} embryo at the beginning of VF formation process (top) and (b) 4 minutes later (bottom): note the variability of apical cell area. (c) Observed changes in edge length $\Delta r_{ij}$, edge orientation angle $\Delta \theta_{ij}$ and myosin level $\Delta m_{ij}$. While edge length shrinks $\sim 75\%$  relative changes in myosin level and edge orientation are considerably smaller. (d-e) Test of compatibility constraint (Eq.5) compares the PDF of the measured $\log\chi$'s (blue) with the control distribution (red) defined by permuting angles. Embryonic mesoderm ({\it d}) exhibits a strong tendency towards compatibility ($\log\chi \approx 0$) while epithelium of the third instar imaginal wing disc ({\it e}), does not. (f) Spatial profile of the isogonal mode amplitude, $\{\Theta_\alpha\}$ (just before invagination) 
describes increasing anisotropic compression of cells towards ventral midline. (g) Fraction of deformation $1-\frac{<|\Delta \bm{r} - \Delta\bm{r}_{iso}|^2>}{<|\Delta \bm{r}|^2>}$ captured by the best-fit isogonal transformation. Each color represents an independent measurement with $~200$ cells. Inset: a graphical comparison of the fit for a single time-point. }
\end{figure}

Returning to the dynamics of VF formation, we
used time-lapse microscopy data on the early stage of ventral furrow formation to examine the deformation of cells as a function of time (see SI for details). 
We found, as shown in Fig. 3(g), that the inferred isogonal modes accounted for about 80\% of the total variance of the dynamic vertex displacement field, which clearly indicates that apical deformation of cells during ventral furrow formation is well approximated by an isogonal transformation. Thus, the cell array appears to behave much like a transiently perturbed ATN, flowing along the isogonal manifold which comprises the set of its (mechanical) equilibrium states (see Fig. 2(b)). Consistent with this interpretation, analyzing time-lapse images of Sqh-GFP we found that cortical-myosin levels do not significantly change during this time despite medial-myosin spiking.  The results not only suggest that an ATN model is suitable for describing tissue behavior, but also provide a simpler set of degrees of freedom that accurately describe the dynamics. The profile of final isogonal deformation $\{\Theta_\alpha\}$ was found to be approximately parabolic (shown in Fig. 3f), consistent with anisotropic constriction of cells with the long axis oriented along the anterior-posterior direction \cite{Wieschaus91}. This could be patterned via graded medial myosin pulses.

\subsection*{Discussion}
ATN model formulated above describes epithelial tissue dynamics in terms of three processes: i) fast relaxation towards mechanical equilibrium dominated by cortical tension; ii) myosin driven rearrangement of cortex on an intermediate time scale and iii) on the slowest timescale, Dynamic Recruitment (or reduction) of myosin driven by the internal rate of strain in the cortex. The 1st two alone would result in a viscoelastic fluid behavior (driven by myosin generated internal forces), Dynamic Recruitment however, dramatically changes the long term behavior so that while being able to flow at short times, ATNs, like solids, can support external stress at long times. 

Tension balance imposes a constraint on cell geometry, which we found to be approximately obeyed in some of the epithelial tissues observed during {\it Drosophila} development.  Existence of isogonal ``soft modes" predicted by the ATN model was strikingly confirmed by the analysis of cellular deformations in the process of Drosophila Ventral Furrow formation, where isogonal modes nicely account for the observed extreme variability of apical area of cells. While these observations confirm the validity of tension balance in describing mechanical equilibrium of epithelial tissue, new experiments will be needed test the Dynamic Recruitment hypothesis, that was introduced to explain how myosin levels at different interfaces can be coordinated to attain equilibrium.

In conclusion, the ATN model describes a very unusual solid - an ``Active Solid" - illustrating the wealth of novel physics associated with Living Matter, while providing insight into biological phenomena.
\subsection*{Materials and Methods}
The following fly stocks where used for ventral furrow live recordings: Spider GFP \cite{Chia02}, sqh-GFP;membrane-mCherry \cite{Martin10}. Embryos where dechoreonated following standard protocols, and mounted in Matek Dishes for imaging. Images where acquired on a Leica SP8 confocal, equipped with a 40x/N.A. 1.1 objective water immersion objective.  See SI for details on image analysis and numerical simulation of ATN dynamics.

\begin{acknowledgments}
The authors gratefully acknowledge stimulating discussions with Ken Irvine, Thomas Lecuit, and Eric Wieschaus and would like to thank K. Irvine for sharing the wing imaginal disc data. This work was supported by the NSF PHY-1220616 (BIS) and by the GBMF grant \#2919 (BIS/IH).
\end{acknowledgments}

\end{document}


\title{Supplementary Information for `Active Tension Network model reveals an exotic mechanical state realized in epithelial tissues'}
\author{Nicholas Noll}
\affiliation{Department of Physics, University of California Santa Barbara}
\author{Madhav Mani} 
\affiliation{Department of Applied Mathematics, Northwestern University }
\affiliation{Kavli Institute for Theoretical Physics}
\author{Idse Heemskerk} 
\affiliation{Department of Biosciences, Rice University}
\affiliation{Kavli Institute for Theoretical Physics}
\author{Sebastian Streichan} 
\affiliation{Kavli Institute for Theoretical Physics}
\author{Boris I. Shraiman}
\affiliation{Department of Physics, University of California Santa Barbara}
\affiliation{Kavli Institute for Theoretical Physics}

\maketitle

\section{Isogonal modes and conformal symmetry.}
Isogonal modes can be thought of as the discretized degrees of freedom associated to the conformal symmetry of the continuum description. The elastic energy associated to displacement field $u_i$ with vanishing bulk modulus is
\begin{equation}
E = \frac{1}{2}\displaystyle\int d^2\bm{r} \left[ \partial_i u_k + \partial_k u_i - \delta_{ik} \partial_l u_l \right]^2
\end{equation}
Assuming relaxational dynamics - i.e. $\dot{u}_i = -\frac{\delta E}{\delta u_i}$  - the equation of motion for field $u_i$ is found to be $\dot{u}_i = \partial^2 u_i$. Any solution of the Cauchy-Riemann equations ($\partial_x w_x =\partial_y w_y ; \ \ \partial_x w_y =-\partial_y w_x$) can be added to $u_i$ with no generation of additional internal stresses. That is to say, any conformal transformation of our equilibrium displacement field is also a valid ground state. Isogonal modes correspond to independent local dilations.

Alternatively isogonal modes can be thought of as generalizations of the Villain's iso-perimetric {\it breather modes} \cite{Villain80} of a hexagonal lattice of domain walls. Villain's model  \cite{Villain80} in particular described adsorbed atoms on a 2D substrate, where, in the incommensurable phase,  `grain' boundaries  form between `out' of register phases. The boundary energy for regular hexagonal lattice of such domains $\sum_{ij} \sigma r_{ij}$ is unchanged breather modes generated by dilations of hexagons, which can be demonstrated to leave the total length of the boundary $\sum_{ij}  r_{ij}$ invariant. Our isogonal modes are a generalization to the case of a general lattice (satisfying ATN equilibrium constraints) when interfacial energy varies from edge to edge $\sigma \rightarrow \sigma_{ij}$.

\section{Mode analysis of the 1D cable}
In the continuum limit, the equations of motion in 1D become
\begin{equation}
\partial_t \begin{pmatrix} \delta r \\ \delta u \\ \delta m \end{pmatrix} = \begin{pmatrix} 0 & \partial_x & 0 \\ 0 &\partial_x^2 - q  & q \\ 0 & \alpha & -\alpha \end{pmatrix} \begin{pmatrix} \delta r \\ \delta u \\ \delta m \end{pmatrix}
\end{equation}
$x$ denotes the coordinate along the cable, expressed in units relative to the lattice spacing. As was expected, all elements of the first column of the matrix are zero, implying that $\delta r$ displacements along the cable are zero modes and that their associated dynamics is slaved to the dynamics of tension and myosin perturbations. Hence, we focus on the reduced myosin/tension system in Fourier space
\begin{equation}
\partial_t \begin{pmatrix} \tilde{u} \\ \tilde{m} \end{pmatrix} = \begin{pmatrix} -k^2-q & q \\ \alpha & -\alpha \end{pmatrix}\begin{pmatrix} \tilde{u} \\ \tilde{m} \end{pmatrix}
\end{equation}
The exact dispersion relation for both branches is 
\begin{equation}
\lambda_{1,2} = -\frac{k^2 + \alpha + q}{2}\left[1 \pm \sqrt{1 - \frac{4\alpha k^2}{(k^2+\alpha+q)^2}}\,\right]
\end{equation}
Assuming mechanical feedback occurs on a slower time-scale than actomyosin contractility ($\frac{\alpha}{q} << 1$) we can expand each dispersion relation to linear order in our small parameter
\begin{equation}
\lambda_{1,2} = -\left[k^2 + q + \frac{\alpha q}{k^2 + q}\right],\,-\frac{\alpha k^2}{k^2 + q} 
\end{equation}
We immediately see that first branch is gapped by $q+\alpha$ while the second branch is acoustic, corresponding to fact that a global rescaling of tension and myosin along the cable ($\delta u = \delta m = $ const.) does not perturb the underlying force balance or stall condition - i.e. there are phonons at long times as it is a solid! The eigenvectors are 
\begin{equation}
\phi_{1,2} = \begin{pmatrix} 1 \\ \frac{-\alpha}{k^2 + q} \end{pmatrix},\,\begin{pmatrix} \frac{q}{k^2 + q} \\ 1 \end{pmatrix} 
\end{equation}
For $\alpha=0$ the gapped mode corresponds solely to tension perturbations. Conversely, for $0<\alpha<<1$, the gapped mode is an admixture between both tension and myosin perturbations along the cable; the myosin component is proportional to $\frac{\alpha}{q}$ and thus small.

We now study the equations under transient forcing on the boundary conditions. We expand tension along the cable in a Fourier sine series
\begin{equation}
T = T_0 + T_\Delta \left(\frac{x}{L}\right) + \displaystyle\sum\limits_{n=1}^\infty \tilde{T}_n \sin\left(\frac{n\pi x}{L}\right)
\end{equation}
We are only interested in symmetric longitudinal pulling and thus set $T_\Delta = 0$. A similar decomposition exists for $m$ which allows us to write the equations
\begin{align}
i\omega \, \bar{T}_n + \left(\frac{n^2\pi^2}{L^2} + q\right) \bar{T}_n - q \bar{m}_n &= \bar{F}_n \\
i\omega \bar{m}_n + \alpha \bar{m}_n - \alpha \bar{T}_n &=\bar{G}_n
\end{align}
The dynamic boundary conditions act as a source as expected (only onto the odd modes as they respect the left/right symmetry). The second equation implies $\bar{m}_n = \frac{\bar{G}_n + \alpha \bar{T}_n}{\alpha + i \omega}$. Substituting into the first equation
\begin{equation}
\bar{T}_n \left[\frac{n^2\pi^2}{L^2} + q\left(1-\frac{\alpha}{\alpha+i\omega}\right) + i \omega \right]  = i\omega \bigg[ \bar{F}_n + \frac{q}{\alpha + i\omega} \bar{G}_n \bigg]
\end{equation}
which can be simplified to obtain (we assume the forcing function on myosin is equivalent to the forcing function on tension)
\begin{equation}
\bar{T}_n =  -\frac{i\omega \left[1 + \frac{q}{\alpha + i\omega}\right] }{\frac{n^2\pi^2}{L^2} + i\omega \left[1+ \frac{q}{\alpha+i\omega}\right] } \frac{(1-(-1)^n)}{\pi n} \bar{F}_{ext}
\end{equation}
Define $b^2(\omega) = i\omega L^2\left[1+\frac{q}{\alpha+i\omega}\right]$ to improve the appearance of the equations. 
\begin{equation}
\bar{T}_n(\omega) = -\frac{b^2(\omega)}{(\pi n)^2 + b^2(\omega)} \frac{(1-(-1)^n)}{\pi n} \bar{F}_{ext}(\omega)
\end{equation}
The series can be re-summed and written in a clean notation shifting $x \in [-L/2,L/2]$. 
\begin{equation}
\bar{T}(x,\omega) = \frac{\cosh(\frac{b x}{L})}{L\cosh(\frac{b}{2})} \bar{F}_{ext} (\omega)
\end{equation}
This immediately implies the phase relationship between strain and the external force is
\begin{equation}
\bar{r}(\omega) = - \frac{i b^2 \cosh(\frac{b x}{L})}{\omega L^2 \cosh(\frac{b}{2})} \bar{F}_{ext} (\omega) 
\end{equation}
We note this has the expected regimes of behavior discussed in the main text (focus on the boundary for simplicity). For $\omega << \alpha$, the relationship is $\bar{r} \sim \left[1+\frac{q}{\alpha}\right] \bar{F}$ and thus it behaves as a spring with stiffness $\frac{\alpha}{q+\alpha}$ as expected from our dispersion relation derived above. For $\alpha << \omega << q$ the relationship is $\bar{r} \sim \frac{i q}{\omega} \bar{F}$ and thus it behaves as a visco-elastic fluid. This regime is where isogonal deformations are expected to exist. Lastly, if $\omega >> q$ then $\bar{r} \sim \bar{F}$ telling us we are pulling on the elastic cytoskeletal network.
\section{Mode analysis of the 2D ATN near equilibrium.}
The dynamics near equilibrium is most naturally expressed in terms of edge vectors $\bm{r}_{\alpha\beta}$, where each edge is now labeled by the cells it partitions, in this case cells $\alpha,\beta$, and is bordered by vertices $(\alpha,\beta,\gamma)$ and $(\beta,\alpha,\gamma')$ (vertices here being labelled by the triples of the adjacent cells. Equations of motion can be derived directly from Eq. $(2)$ in the main text. Hereafter time is rescaled $t \to \frac{\kappa}{\nu} t$ to reduce the appearance of unnecessary constants. 
\begin{equation}
\frac{d}{dt} \bm{r}_{\alpha\beta} = \bm{u}_{\beta\gamma} + \bm{u}_{\gamma\alpha} + \bm{u}_{\beta\gamma'} + \bm{u}_{\gamma'\alpha} - 2 \bm{u}_{\alpha\beta}
\end{equation}
where $\bm{u}_{\alpha\beta} \equiv \kappa^{-1} T_{\alpha\beta} \bm{\hat{r}}_{\alpha\beta} = u_{\alpha\beta}  \bm{\hat{r}}_{\alpha\beta}$. Parameterization in terms of edge vectors simplifies the resultant algebra at the cost of introducing $2c$ additional degrees of freedom associated to the geometric constraint that edge vectors sum to zero around each cellular plaquette 
\begin{equation}
\displaystyle\sum\limits_{\{\beta\}_\alpha} \bm{r}_{\alpha\beta} = 0 \quad \forall \alpha
\end{equation}
It is easy to check that dynamics described by Eq. $(15)$ preserves the constraint defined by Eq. $(16)$. We linearize Eq. $(15)$ and decompose into transverse $\delta \theta_{\alpha\beta}$ and longitudinal $\delta r_{\alpha\beta}$ modes defined by
\begin{equation}
\delta \bm{r}_{\alpha\beta} = \delta r_{\alpha\beta} \, \bm{\hat{r}}_{\alpha\beta} + \delta \theta_{\alpha\beta} \, r_{\alpha\beta} ( \bm{\hat{z}} \wedge \bm{\hat{r}}_{\alpha\beta} )
\end{equation}
leaving us with equations
\begin{align}
\frac{d}{dt} \delta r_{\alpha\beta} &= \sum_{\gamma\gamma'} \left [ L_{\alpha\beta;\gamma\gamma'} \delta u_{\gamma\gamma'} - A_{\alpha\beta;\gamma\gamma'} \delta \theta_{\gamma\gamma'} \right ] \\
r_{\alpha\beta} \frac{d}{dt} \delta \theta_{\alpha\beta} &= \sum_{\gamma\gamma'} \left [ A_{\alpha\beta;\gamma\gamma'} \delta u_{\gamma\gamma'} + L_{\alpha\beta;\gamma\gamma'} u_{\gamma\gamma'} \delta \theta_{\gamma\gamma'}\right ]
\end{align}
where we have defined
\begin{align*}
L_{\alpha\beta;\gamma\gamma'} &\equiv \bm{\hat{r}}_{\alpha\beta}\cdot\bm{\hat{r}}_{\gamma\gamma'} \left[\delta_{\beta\gamma} - \delta_{\alpha\gamma} + \delta_{\alpha\gamma'} - \delta_{\beta\gamma'}\right]  \\
A_{\alpha\beta;\gamma\gamma'} &\equiv \bm{\hat{r}}_{\alpha\beta}\wedge\bm{\hat{r}}_{\gamma\gamma'} \left[\delta_{\beta\gamma} - \delta_{\alpha\gamma} + \delta_{\alpha\gamma'} - \delta_{\beta\gamma'}\right]
\end{align*}
Dynamics of small perturbations in intrinsic length is found by expanding Eq. $(3)$ from the main text about the fixed point
\begin{align}
\frac{d}{dt} \delta \ell_{\alpha \beta } \, &= \, q_{\alpha \beta} \left[ \delta u_{\alpha \beta } - \delta m_{\alpha \beta } \right]
\end{align} 
where 
\begin{equation}
q_{\alpha\beta} \equiv  \frac{\nu \ell_{\alpha\beta}}{\kappa u_{\alpha\beta}} W'(1)
\end{equation}
Tension dynamics is easily obtained via the constitutive relation $u_{ij} = r_{ij} - \ell_{ij}$
\begin{align}
\frac{d}{dt} \delta u_{\alpha \beta } = \frac{d}{dt} \delta r_{\alpha\beta} -q_{\alpha \beta} \delta u_{\alpha\beta} + q_{\alpha\beta} \delta m_{\alpha\beta}
\end{align} 
Lastly, the myosin dynamics is governed by
\begin{equation}
\frac{d}{dt} \delta m_{\alpha\beta} = \alpha \left( \delta u_{\alpha\beta} - \delta m_{\alpha\beta} \right)
\end{equation}
where myosin has been rescaled to have units of interfacial deformation: $\delta m_{\alpha\beta} \to T_s \kappa^{-1} \delta m_{\alpha\beta}$ and $\alpha \equiv \mu \nu \kappa^{-1} W'(1)$. Isogonal modes correspond to $\delta \theta =\delta u =0$ which is realized by $\delta \ell_{\alpha \beta } =\delta r_{\alpha \beta } $, provided $\sum_{\{ \beta \}_{\alpha}} \delta r_{\alpha \beta } \bm{{\hat r}}_{\alpha \beta} =0$. The latter constraint is satisfied for 
\begin{equation}
\delta \bm {r}_{\alpha \beta \gamma}=\hat{\bm r}_{\alpha \beta} \frac{T_{\alpha \beta} \Theta_{\gamma}}{S_{\alpha\beta\gamma}}+\hat{\bm r}_{ \beta \gamma} \frac{T_{\beta \gamma} \Theta_{\alpha}}{S_{\alpha\beta\gamma}}+
\hat{\bm r}_{\gamma \alpha} \frac{T_{\gamma \alpha} \Theta_{\beta}}{S_{\alpha\beta\gamma}}
\end{equation}
where $\delta \bm {r}_{\alpha \beta \gamma}$ denotes displacement of vertex at which adjacent cells $\alpha, \beta, \gamma$ meet; $\Theta_{\alpha},\Theta_{\beta}, \Theta_{\gamma}$ are independent variables associated with these cells and $S_{\alpha\beta\gamma}$ denotes the area of said vertex's dual triangular plaquette. Thus, isogonal deformations are parameterized by $\{\Theta_{\alpha}\}$ and have no restoring force.

Eqs. $(18-19)$ and $(22-23)$ fully specify the closed form linearized dynamics with matrix $H$ that can be expressed  
\begin{equation}
S = U^{-1}HU
\end{equation}
where $U \equiv \text{diag}\left[1,1,\sqrt{u_{\alpha\beta} r_{\alpha\beta}},\sqrt{\frac{q_{\alpha\beta}}{\omega}}\right]$ and
\begin{align*}
S\equiv 
\begin{pmatrix} 0 & L_{\alpha\beta;\gamma\gamma'} & -A_{\alpha\beta;\gamma\gamma'} & 0 \\ 0 & L_{\alpha\beta;\gamma\gamma'} - q_{\alpha\beta}\delta_{\alpha\beta;\gamma\gamma'} & -A_{\alpha\beta;\gamma\gamma'}\sqrt{\frac{u_{\gamma\gamma'}}{r_{\gamma\gamma'}}} &  \sqrt{\alpha q_{\alpha\beta}}\delta_{\alpha\beta;\gamma\gamma'} \\ 0 & \sqrt{\frac{u_{\alpha\beta}}{r_{\alpha\beta}}}A_{\alpha\beta;\gamma\gamma'} & \sqrt{\frac{u_{\alpha\beta}}{r_{\alpha\beta}}}L_{\alpha\beta;\gamma\gamma'}\sqrt{\frac{u_{\gamma\gamma'}}{r_{\gamma\gamma'}}}& 0 \\ 0 & \sqrt{\alpha q_{\alpha\beta}} \delta_{\alpha\beta;\gamma\gamma'} & 0  & -\alpha\delta_{\alpha\beta;\gamma\gamma'} \end{pmatrix}
\end{align*}
Because the first column of the matrix is equal to zero, $\delta r_{\alpha \beta}$ is slaved to other components and thus the rank of $H$ is at most $9c$ as our null space contains $c$ isogonal modes defined above, along with the $2c$ geometric constraints (see Eq. $(16)$) that are conserved by the dynamics. The left eigenvectors of isogonal modes were numerically found to be exponentially localized around the respective cell with a length scale $q^{-1/2}$: i.e. they are only forced with the screening length set by contractility as shown in Fig. 2 (a). 

The reduced matrix $\tilde{S}$ is obtained by eliminating the $1^{st}$ row and $1^{st}$ column of $S$. It is manifestly symmetric in our chosen basis, following immediately from the fact that $L_{\alpha\beta;\gamma\gamma'}$ and $A_{\alpha\beta;\gamma\gamma'}$ are symmetric and anti-symmetric respectively. Furthermore, it is easy to see that $L_{\alpha\beta;\gamma\gamma'}$ satisfies all properties of a   normalized weighted graph Laplacian defined over edges in our triangulation and thus will be negative semi-definite, as shown in Fig. 1, ensuring stability of the unperturbed ATN state
\begin{figure}[h]
\centerline{
\includegraphics[width=.5\textwidth]{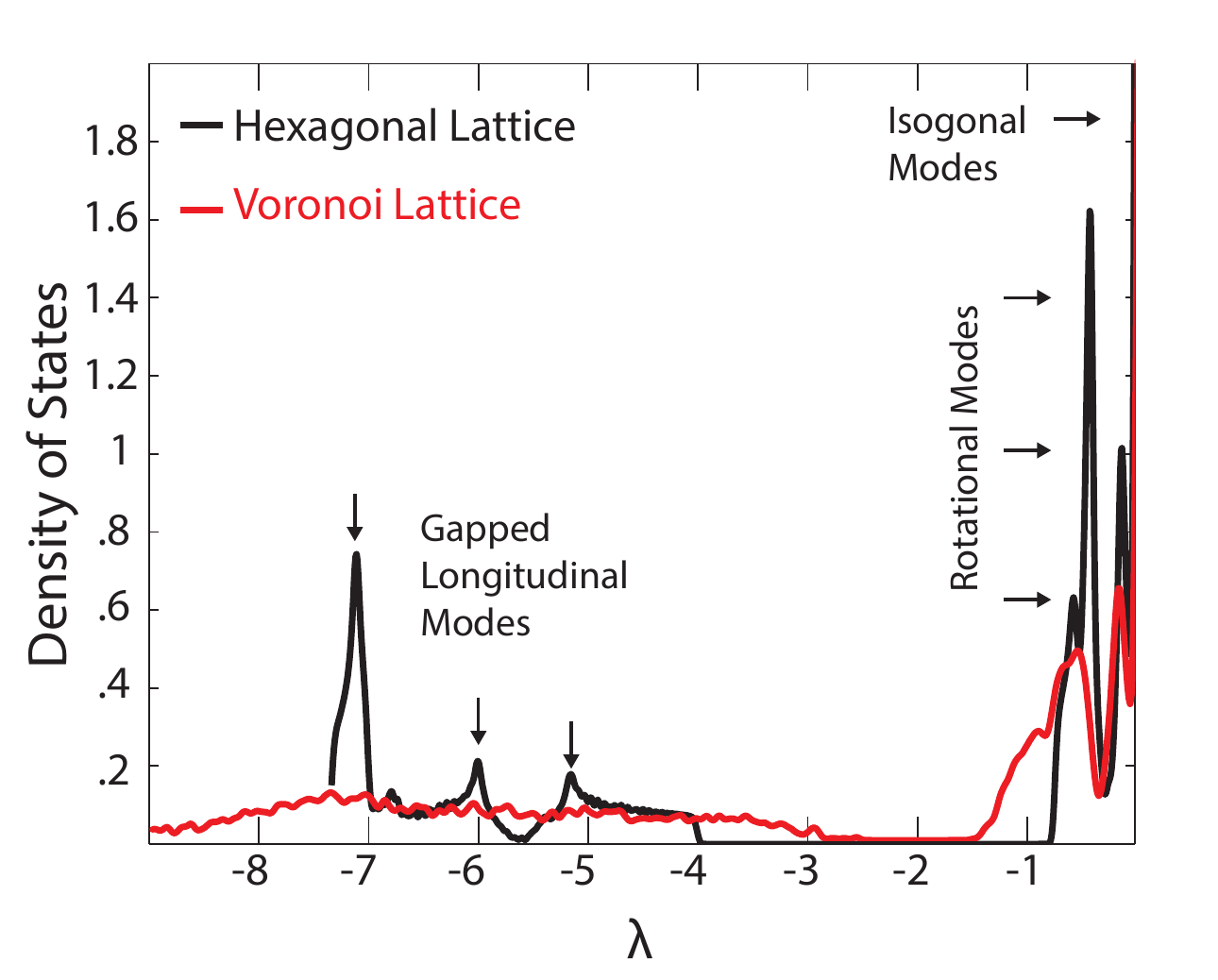}
}
\caption{``Density of states" plot for the normal modes of $H$ governing the dynamics of fluctuations about the ATN equilibrium corresponding to i) a hexagonal array (black line) and ii) a randomly generated Voronoi tesselation (red line). In both cases  $3c$ modes lie at zero, of which $c$ are the isogonal modes and the remaining $2c$ correspond to geometric constrains (on edge vectors). The rest of the eigenvalues are negative as required by stability. $1/3$ of the modes (in the hexagonal lattice case) are separated from zero by a gap proportional to the activity parameter $q$.}
\end{figure}
Another important characterization of the normal modes is the structure of eigenmodes: are they localized or extended? To address this question we numerically measured the distribution of participation ratios, defined as
\begin{equation}
p_\nu \equiv \displaystyle\sum\limits_{m=1}^{N} |\phi_m^\nu |^4
\end{equation}
(where $\nu$ labels the eigenmode) as a function of system size $N$. If $\phi^\nu$ is extended, then $|\phi_m^\nu| \sim 1/\sqrt{N}$ and thus $p_\nu$ should scale with inverse system size. Similarly, if $\phi^\nu$ is localized, it should saturate to a finite number with increasing $N$. We tested the localization of our modes by tracking how the distribution of $p_\nu$ scaled with increasing number of cells within hexagonal and randomly generated voronoi lattices. Isogonal modes were excluded from analysis as it is known a priori that each is localized to a single cell. All non-isogonal modes are fully extended in the hexagonal case - the system is diagonalizable in a plane-wave basis - as shown numerically in Fig. 2(b). Conversely, as shown in Fig. 2(c), it was found that all but one band of `transverse' modes localize for disordered Voronoi lattices. In other words, $c$ modes are still fully extended on a disordered triangulation. 

\begin{figure}[h]
\centerline{
\includegraphics[width=1\textwidth]{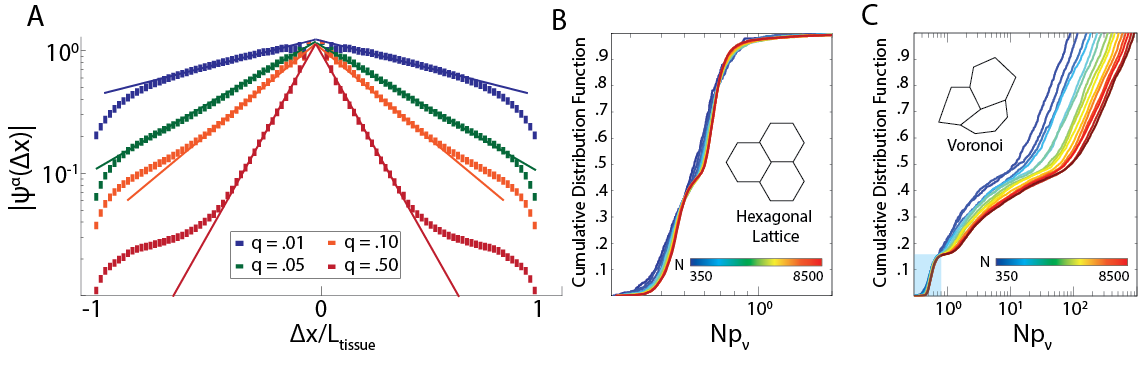}
}
\caption{(a) Eigenvectors associated to isogonal modes fall off exponentially with the characteristic length scale of decay $\sim q^{-1/2}$. Points correspond to left eigenvectors obtained by numerical diagonalization of $H$; solid lines are decaying exponentials with characteristic length $q^{-1/2}$. (b) Eigenmode structure for regular hexagonal and voronoi lattices: fraction of eigenmodes (excluding isogonal modes) below a given participation ratio $p_\nu N$ (defined by Eq. $26$))  scaled with the number of cells $N$. All modes are extended as indicated by the collapse of curves scale on top of each other. (c)  Repeated for random Voronoi lattices of varying size. In contrast to the case (b), only 1/7 of the modes (highlighted in light blue) are extended, while $6/7$ are localized, as indicated by the lack of curve collapse.}
\end{figure}
\section{Validity of the empirical null distribution of $\chi$ and further tests}
\begin{figure}[H]
\centerline{
\includegraphics[width=.75\textwidth]{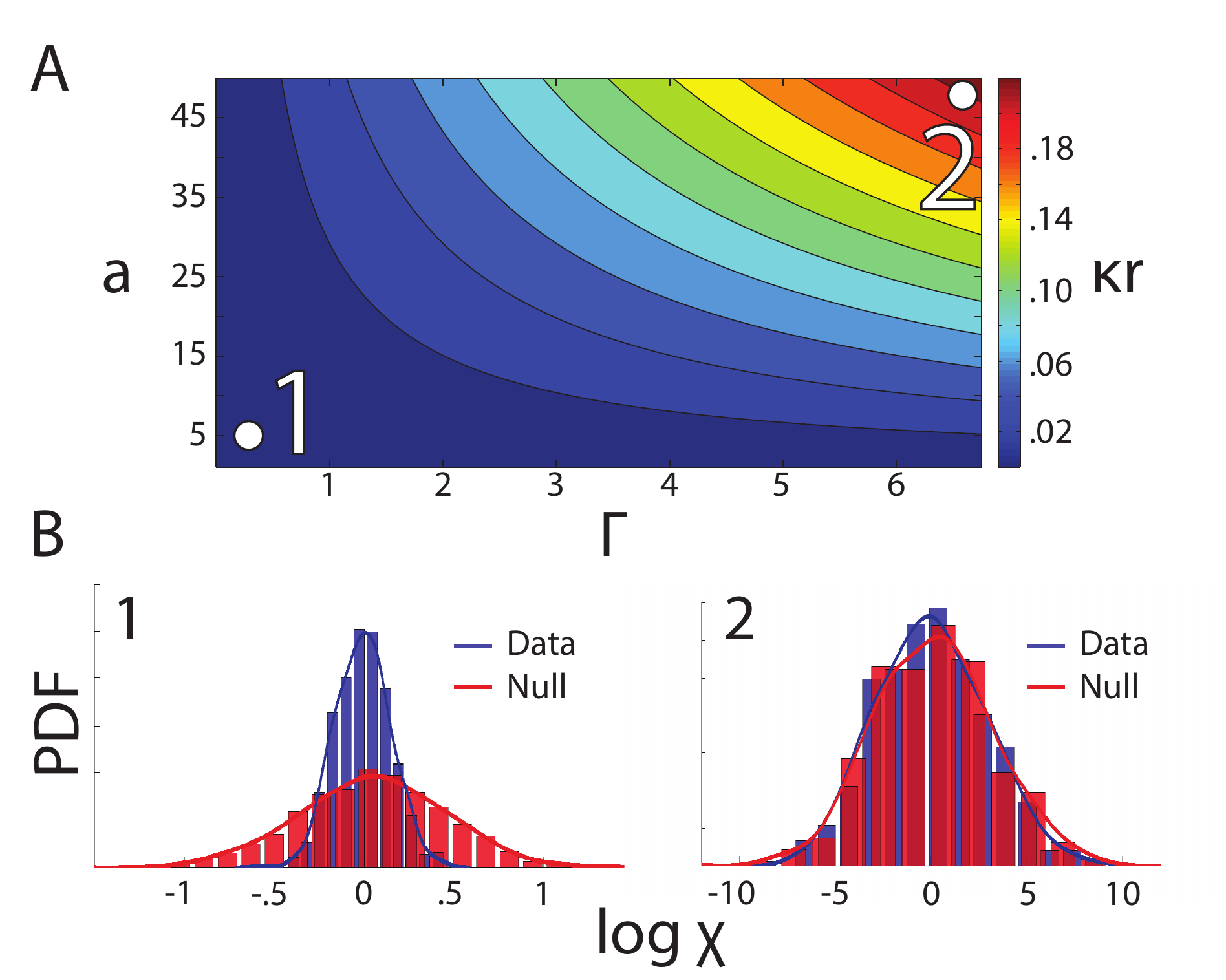}
}
\caption{(a) A contour plot - in the space of vertex model parameters, $a$ and $ \Gamma$, defined by Eq. (1) - of average edge curvature in units of the edge length, $Kr$. In a static tension net cell edges are straight hence curvature would be expected to be exactly zero and is thus found in the lower left corner of the plot. Conversely, as the area parameters become large, the lattice approaches a regime where pressure can no longer be neglected, found in the upper right corner.  (b) Two histograms of the compatibility condition measured from the synthetic cell arrays. Numbers match the numbered points in (a). As $\Gamma \to 0$ we approach our static tension net limit. As expected, for parameters corresponding to point 1, the vertex-model generated synthetic tissue exhibits a compatibility measure, $\log \chi$, tightly clustered around zero relative to the null distribution. Parameters for point 2 are in the regime where pressure differentials between cells are important and tension network approximation is not valid: in this case disrtibution of $\log \chi$ is statistically consistent with the empirical null. }
\end{figure}
To motivate our model, we compared empirically measured distributions of the compatibility condition to a `random' cell array constructed from the tissue's measured angle distribution. In other words, we construct our `test' cell array by building cells with interior angles sampled from the distribution of all angles within the entire array, resulting in a cell array that won't obey compatibility by construction. This null distribution can be used as a baseline to which our empirical distribution can be compared against; if the `true' distribution looks identical to our null distribution then we are forced to reject the `tension-net' hypothesis. The outlined procedure was checked against synthetic data to test its validity. Cell arrays of approximately 100 cells were relaxed to their equilibrium configuration under the energy functional
\begin{equation}
E = \displaystyle\sum\limits_{<i,j>} (r_{ij} - 1)^2 + \Gamma\displaystyle\sum\limits_{\alpha} (A_{\alpha} - a)^2 
\end{equation}
The quantity $\langle\kappa r \rangle = \frac{\langle \Delta P r\rangle}{\langle T \rangle}$ was measured for different values of $\Gamma$ and $a$. The resultant contour plot is displayed in Fig. 2(a).

Once the cell arrays are relaxed, one can `pixelate' the tissue and then measure the distribution for the compatibility condition $\chi$, c.f. Eq. 5 in the main text, and compare against the null construction in the exact same manner as was done on actual data. As is shown in Fig. 3(b), close to the static tension net limit (point 1), the empirical compatibility distribution clusters much closer to zero than the null. Conversely, when $K r \sim .2$ (corresponding to point 2), the data resolves exactly the same as our null distribution showing a failure of the `static tension hypothesis.' 

Our statistical test was tested against four different epithelial tissues, two of which were discussed in the main text. The full set is shown below: ventral ectoderm minutes before ventral furrow formation, pupal notum, lateral ectoderm during early germ band extension, and third instar larval imaginal wing disc (data kindly provided by Ken Irvine). Proprietary segmentation code was used to process all live image movies into vertex model `skeletonizations' over time. Example images of analyzed data are shown below. 
\begin{figure}[H]
\centerline{
\includegraphics[width=\textwidth]{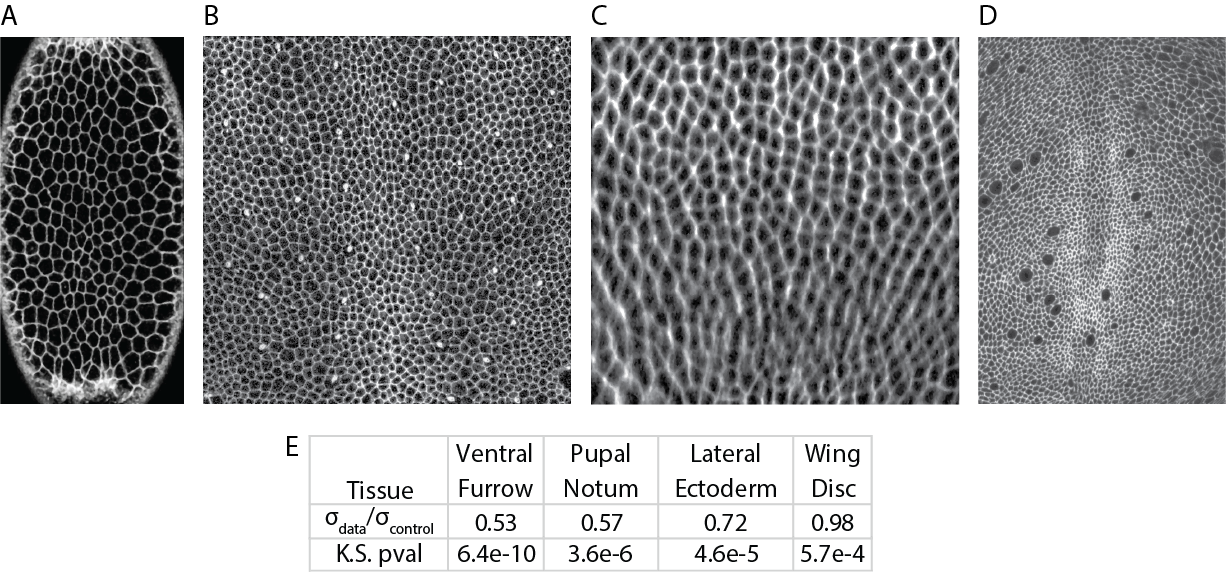}
}
\caption{(a) An image of the ventral ectoderm ({\it Drosophila} embryo) minutes before invagination of the ventral furrow; (b) Pupal notum epithelium; (c) Embryonic lateral ectoderm during early germ band extension; (d) Epithelium of the wing imaginal disc at the third instar larval stage. (e) Characterization of the distribution of the compatibility measures ($\log \chi$ see Eq. (9) in the main text) for the four tissues. Comparing the standard deviations of the measured distribution to that for the control distribution: the closer to the zero their ratio $\sigma_{data}/\sigma_{control}$  is, the closer is the observed lattice is to a true static tension net. Statistical significance of the difference between the distributions is demonstrated by the Kolmogorov-Smirnov p-value given in the second row of the table.}
\end{figure}
\section{Procedure used to fit Isogonal Deformation during Ventral Furrow Formation}
The fundamental equation to invert to measure `isogonal' deformation is Eqn. (6) in the main text, reproduced here
\begin{equation}
\delta \bm {r}_{\alpha \beta \gamma}=\hat{\bm r}_{\alpha \beta} \frac{T_{\alpha \beta} \Theta_{\gamma}}{S_{\alpha\beta\gamma}}+\hat{\bm r}_{ \beta \gamma} \frac{T_{\beta \gamma} \Theta_{\alpha}}{S_{\alpha\beta\gamma}}+
\hat{\bm r}_{\gamma \alpha} \frac{T_{\gamma \alpha} \Theta_{\beta}}{S_{\alpha\beta\gamma}}
\end{equation}
This immediately introduces two problems: (i) we must track vertices over time to measure the deformation field $\delta \bm{r}_{\alpha\beta\gamma}$  and (ii) we must be find a `close' exactly compatible cell array in order to be able to isogonally dilate and contract cells. The latter is equivalent to `inferring' the underlying tension triangulation for a given cell array that will be described in a future write-up. Once tensions are found, the geometric factors of the triangulation directly enter the matrix entries defined by Eqn. $(28)$. Predicted tensions as well as the cumulative distribution is shown in Fig. 5 (ab) respectively for a snapshot of ventral furrow formation.
\begin{figure}[H]
\centerline{
\includegraphics[width=\textwidth]{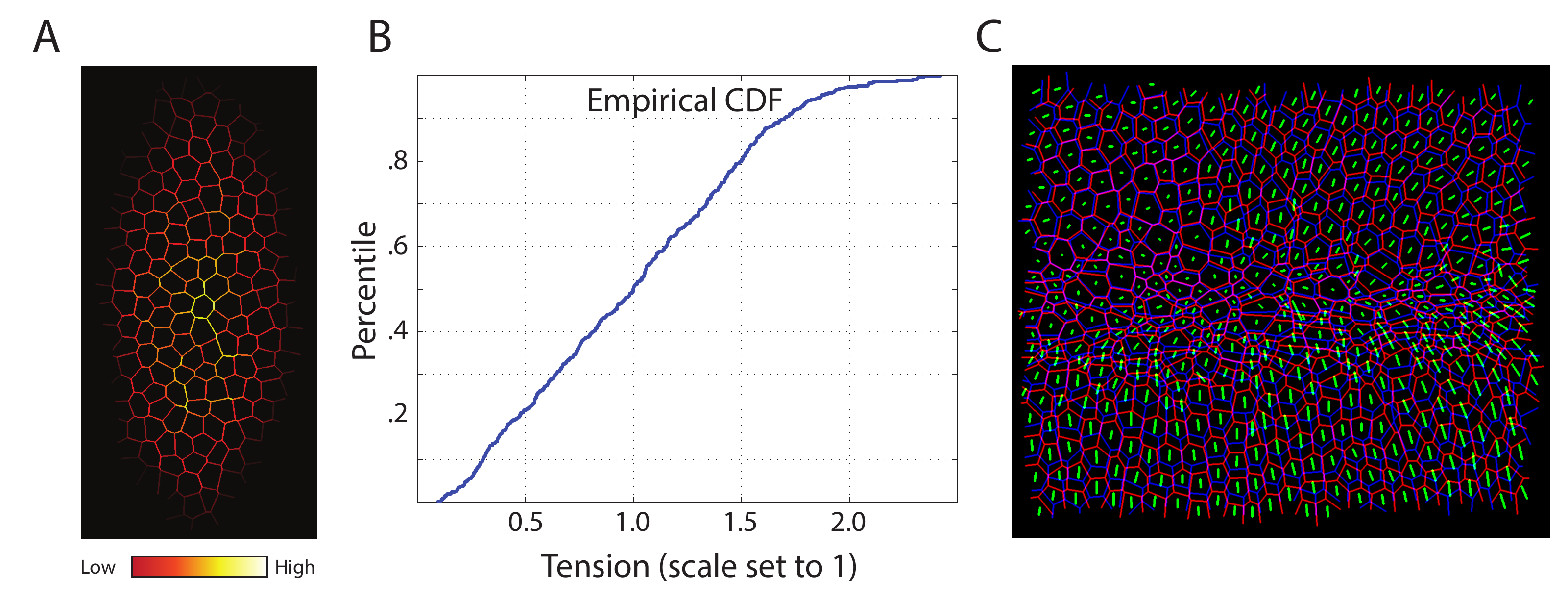}
}
\caption{(a) A heatmap illustrating the distribution of tensions on the ventral side of the embryo minutes before ventral furrow formation. The `hotter' the color, the higher the tension. (b) A cumulative distribution function of all estimated balanced tensions within the cell array shown in (a). Due to the assumption of force balance the scale is unknown and thus the mean tension is set to 1. The distribution is relatively uniform. (c) Example of cell tracking algorithm pairing segmented cells in subsequent time-points.}
\end{figure}
Vertices were tracked by tracking cells' via pixel overlaps and using tracked cells to define vertex displacements $\delta \bm{r}_{\alpha\beta\gamma}$ for successive time points (20 second intervals for all movies analyzed) by looking for vertices that share the same three bordering cells. These displacements were directly used on the L.H.S. of Eqn. (28).

Eqn. $(28)$ is a rectangular ($2v$ by $c$) linear system of equations defining vertex displacements corresponding to an arbitrary isogonal transformation parameterized by $\{\Theta_\alpha\}$ and can be solved by simple least squares analyses.  This problem is heavily over-constrained ($2v = 4c$ as compared to $c$ fitting parameters) and thus represents a strong test of our proposed ATN model. 

\section{Image Analysis Methods}
All images were first classified using machine learning software Ilastik \cite{Sommer11}. The resultant probability map was passed into MATLAB and segmented using the watershed algorithm \cite{Meyer94} after pre-filtering. Once segmented, all relevant quantities such as vertex position and neighboring cells and bonds were stored in a custom data structure. All code is available upon request.

Tracking (matching segmentation labels between subsequent time-points) was done using point-matching of cell centroids after correcting for PIV (Particle Image Velocimetry) estimated flow fields between time points. PIV flow fields were estimated using cross-correlation between gridded regions defined on our image \cite{Adrian92} With cell's tracked, vertices and bonds can be easily tracked using their bordering cells. 

\section{Simulation Methods}
Eqs. (2-4) in the main text were numerically integrated using MATLAB's ODE15s solver as the time-scale separation resulted in a stiff system. T1 events were handled using MATLAB's event feature, if an edge falls below a critical user-specified value, then we flag an event which stops the integration. A T1 event is manually performed and then numerical integration is restarted. 

For the 2D rheology simulation, a 15 x 15 square of cells was initialized in a slightly disordered hexagonal lattice under constant pressure to balance against the internal tension. Sinusoidal external forces were attached to the vertices on the vertical boundary. Strain rate was measured on vertical junctions throughout the bulk.